# Ethics in conversation

Building an ethics assurance case for autonomous AI-enabled voice agents in healthcare

Marten H. L. Kaas

University of York, marten.kaas@york.ac.uk

Zoe Porter

University of York, zoe.porter@york.ac.uk

Ernest Lim

Ufonia Limited, el@ufonia.co

Aisling Higham

Ufonia Limited, ah@ufonia.co

Sarah Khavandi

Ufonia Limited, sk@ufonia.co

Ibrahim Habli

University of York, ibrahim.habli@york.ac.uk

The deployment and use of AI systems should be both safe and broadly ethically acceptable. The principles-based ethics assurance argument pattern is one proposal in the AI ethics landscape that seeks to support and achieve that aim. The purpose of this argument pattern or framework is to structure reasoning about, and to communicate and foster confidence in, the ethical acceptability of uses of specific real-world AI systems in complex socio-technical contexts. This paper presents the interim findings of a case study applying this ethics assurance framework to the use of Dora, an AI-based telemedicine system, to assess its viability and usefulness as an approach. The case study process to date has revealed some of the positive ethical impacts of the Dora platform, as well as unexpected insights and areas to prioritise for evaluation, such as risks to the frontline clinician, particularly in respect of clinician autonomy. The ethics assurance argument pattern offers a practical framework not just for identifying issues to be addressed, but also to start to construct solutions in the form of adjustments to the distribution of benefits, risks and constraints on human autonomy that could reduce ethical disparities across affected stakeholders. Though many challenges remain, this research represents a step in the direction towards the development and use of safe and ethically acceptable AI systems and, ideally, a shift towards more comprehensive and inclusive evaluations of AI systems in general.

## 1 INTRODUCTION

As AI-based systems increasingly permeate society, it is widely recognized that new approaches to ensuring the safety and efficacy of such systems are needed. But merely ensuring the safety of AI-based systems is not enough. The human tendency to defer to suggestions generated by AI systems, their "black box" and dynamically updating nature, gaps in regulation and an emphasis on being first to market all conspire to threaten not just the safe deployment and use of AI systems, but their ethical acceptability as well. This paper attempts to address the gap between meeting minimum safety requirements and ethical acceptability by evaluating the plausibility, viability and value of instantiating the ethics assurance argument pattern proposed by Porter et al. [41] in the healthcare context for an AI-based telemedicine system. Our interest is not only in safety, but rather something more ambitious: ethical acceptability. As impressive as AI systems are, their abilities are still derived from humans and as such lack the sort of normative commitments and capacity for considered judgement that humans have [47]. It therefore falls on us, the developers, investors, regulators, users, researchers and affected stakeholders, to carefully consider the consequences of deploying AI systems. Our research is, we maintain, one step towards ensuring the responsible development of AI systems whose impacts can be difficult to predict, far-reaching and long lasting.

This paper is structured as follows. In section 2, we introduce the system, Dora, and describe its place in the clinical pathway as well as the regulatory landscape governing its use. In section 3, we describe the principles-based ethics assurance argument pattern and in section 4 apply the argument pattern to Dora and explain our preliminary results. Lastly, in section 5 we draw out some conclusions of our research including limitations of our work and areas for future research.

## 2 THE TECHNOLOGY (DORA) AND ITS CONTEXT

### 2.1 Introduction to Dora

Healthcare is facing a workforce crisis. In the UK, demand on the National Health Service (NHS) is increasing beyond the current capacity of healthcare staff [50]. With increasing demands, and a shortage of healthcare workers, new ways of working with artificial intelligence (AI) enabled tools can help us to meet demand. These tools range from clinical decision support systems (CDSS) [23], to AI-enabled clinical services that make fully autonomous diagnoses such as radiographic imaging systems [4].

The clinical conversation is a fundamental part of the delivery of healthcare. This is an area in which AI-enabled automation can offer significant relief to an over-stretched healthcare system and workforce.

One example of meeting this need is 'Dora', an autonomous, voice-based, natural-language clinical assistant that has clinical consultations with patients over the telephone [27]. Dora speaks to patients via a phone call, meaning there is no requirement for patients to have access to or experience with any 'digital technologies'; they simply receive a phone call on their landline or mobile phone and speak to the system naturally. The aim of the system is to provide a like-for-like replacement of routine clinical calls, whilst identifying patients with symptoms or complications which mean they need to speak to a human clinician. Dora is currently in routine use in the NHS across a number of clinical pathways, and is commissioned across 12 Trusts.

Technically, Dora incorporates a number of different AI technologies. During phone calls, streamed audio from the patient is converted to text input in real-time using a blend of commercial application programming interface-based (API) services to optimise the transcription. To classify the patient's conversation inputs, natural language processing is conducted on the text using a custom entity and intent extraction pipeline [19]. Custom tooling allows continual testing and training to take place to improve the system's capacity to deal with a broad spectrum of inputs from diverse patient's clinical conversations.

To respond to the patient's inputs, a conversation machine learning model, which has been trained on complete conversation flows to enable contextual conversations and deliver secure modular dialogue, is used. Finally, the output from the conversation engine is converted to audio using a commercial API and is streamed back to the patient on the call. These different AI processes are conducted in real-time without noticeable latency. All of the data associated with the call is stored securely in local data-centres. The developer acts as the data processor with the individual hospital client remaining the data controller.

The Dora platform can conduct a variety of clinical conversations, all of which are part of high volume-low complexity clinical pathways. The most evidenced and widely deployed pathway is the cataract pathway. Pre-publication results from a study looking at safety and efficiency of using Dora for post-operative cataract follow-up shows a high sensitivity 94% (95% CI: 85-98) and specificity of 86% (95% CI:80-92) when comparing Dora's clinical recommendation to that of an ophthalmologist blinded to Dora's decision [39]. More recently, in a real-world setting, 1015 consecutive Dora calls were retrospectively reviewed, demonstrating that, of the 742 patients that answered and completed the call, 445 (60%) no longer required a clinician appointment after assessment with Dora [22]. There is also evidence that the Dora technology is acceptable to patients as part of their care pathway. In a study focused on acceptability, 170 patients with a mean age of 76 gave Dora a median net promoter score of 9 out of 10, and high rates of acceptability were documented using the Telephone Usability Questionnaire [27].

### 2.2 How Dora fits into the clinical pathway

Like many digital technologies, the full benefit of Dora can only be realised when it is effectively integrated into hospital clinical pathways. At any clinical site, before calls are delivered, information governance and clinical safety approval are sought. A standard operating procedure (SOP) is agreed with each clinical team, usually with input from the lead clinician, manager, and representatives from each clinical area where Dora will be implemented.

The standard model describing how Dora calls fit into a pathway is shown in Figure 1. Generally, patients are booked into a Dora clinic by admin staff in the same way they are booked for a clinician-led appointment. A list of patients is then sent to the developer (either via email or integration) who generates the Dora calls. After the calls are autonomously delivered by Dora, they are quality assured by the developer. The outcomes from the calls are then returned to the hospital either via email or through electronic healthcare record integration as per the SOP. For some conversations, a clinician reviews all the outcomes and then arranges additional clinical consultation for those patients in whom Dora identified potential concerns.

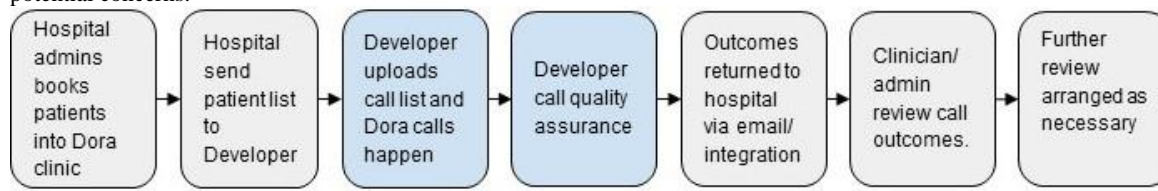

Figure 1: A typical illustration for how a Dora call fits into the pathway from a healthcare provider's perspective.

### 2.3 The regulatory context

The pace of development of AI technologies underlying increasingly autonomous healthcare systems is rapid. For example, state-of-the-art natural-language AI models such as OpenAI's ChatGPT and Google's Med-PaLM have recently even demonstrated the ability to pass medical school exams with minimal additional training [28,46]. Similarly, an increasing number of algorithms seem to exceed human performance for tasks such as the reporting of a plethora of images, scans and investigations across various specialties and modalities [4,8,10].

Despite this rapid development and regulatory approval for these technologies in healthcare in the UK and beyond, there are concerns around current frameworks for ensuring the effectiveness and trustworthiness of these technologies in a real-world clinical setting. Questions remain around how to analyse, report, and act upon errors and potential harms from their real-world use, whilst having an agile and capable regulatory framework for addressing their responsible deployment through product life-cycles [45]. Regulators and academics [29] are responding, for example, with the Food and Drug Administration's (FDA) AI/machine learning (ML) software as a medical device action plan [12] and the Medicines & Healthcare Products Regulatory Agency's (MHRA) change programme for regulating software and AI as a medical device [33], but gaps exist in our understanding of how we maintain the trustworthiness of these ever-learning systems as we scale their adoption across new clinical settings and incorporate the latest AI models (Figure 2). Additionally, it is increasingly clear that a gap exists between legislated minimum safety requirements and ethical acceptability.

The Dora platform is a medical device Class 1 as an active medical device (clinical management support software), under Directive 93/42/EEC, classification rule 12. The product has a registered UKCA mark Class 1 for its intended use as recognised by the MHRA. In most deployments, its use is also subject to NHS England standards DCB0129 and DCB0160 [35]. These two standards are designed to help manufacturers of health information technology (IT) software evidence the clinical safety of their products and healthcare organisations assure the clinical safety of their health IT software, respectively.

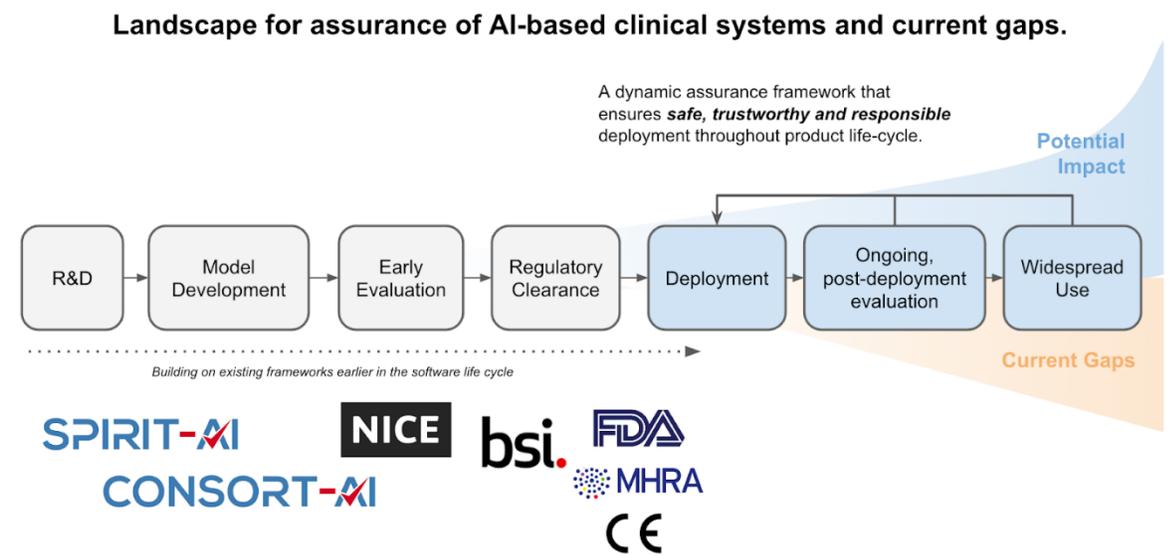

Figure 2: An overview of best-practice standards and guidance for assurance of AI-based clinical systems.

As systems like Dora move beyond meeting the bar for initial regulatory clearance and clinical safety into widespread use, few practical frameworks exist to help guide developers and users towards 'what good looks like' when it comes to continual model monitoring, maintenance, and update. The post-deployment phase of development in the product life cycle is also where clinical AI systems have the most potential for exponential patient and system impact, but also where existing guidance has struggled to provide useful and comprehensive guidance around how to responsibly assure these systems at scale [7,13]. Tension exists between realising the full potential of AI models by allowing them to 'learn' from real-world input data and improve dynamically, but also maintaining the high levels of assurance and trust that the systems continue to perform at the level that they did following regulatory clearance.

Responsible Research and Innovation (RRI) policies encourage actors and institutions to go beyond regulatory compliance and strive for ethically acceptable and ethically sustainable science, technology and innovation outcomes [48,49]. If current regulation and standards set a 'minimum threshold' for the acceptability of systems like Dora, the principles-based ethics assurance argument offers a higher, more comprehensive and more ambitious threshold. It can be seen as an example of RRI in action: a practical, dynamic, and human-centred assurance framework to provide guidance on how AI systems can scale and deploy to wide use whilst maintaining high levels of performance and trustworthiness to clinicians, patients and the public.

## 3 THE PRINCIPLES-BASED ETHICS ASSURANCE ARGUMENT

### 3.1 Overview of the argument

In recent years, the recognition that, while AI can bring benefits, the use of AI can also cause a wide range of harms and entrench existing inequalities has led to a proliferation of ethics declarations and ethical principles for AI [15,24]. Several researchers have noticed a striking overlap between the recurring themes in these declarations and the four classical principles of biomedical ethics [3]: beneficence (bringing benefit); non-maleficence (preventing harm); respect for human autonomy; and justice [16,17,34]. Though these principles are most closely associated with medical ethics, they are not limited to a medical AI context. Another common theme in sets of AI ethics principles is the emphasis on transparency [15,24].

The principles-based ethics assurance argument pattern [41] offers a route to translating these principles into a practicable, human-centred assurance framework. It combines the four ethical principles (adapted to the AI context) and a supporting principle of transparency (the '4+1 ethical principles') with the assurance case methodology. This is a methodology for presenting a clear, structured and defensible argument that justifies confidence in a desired goal, concerning a property of interest [26]. Within engineering, that property of interest has most commonly been safety. Ethics assurance covers a broader range of normative properties of interest [5].

There are several different notations for presenting assurance cases; the principles-based ethics assurance argument pattern uses the Goal-Structuring Notation (GSN) [26,31]. Assurance cases presented in GSN are hierarchically decomposed: there is a top-level goal which is supported, via an argument strategy, by sub-goals, which in turn are supported by evidence. The modular structure of the argument is provided in Figure 3 below.

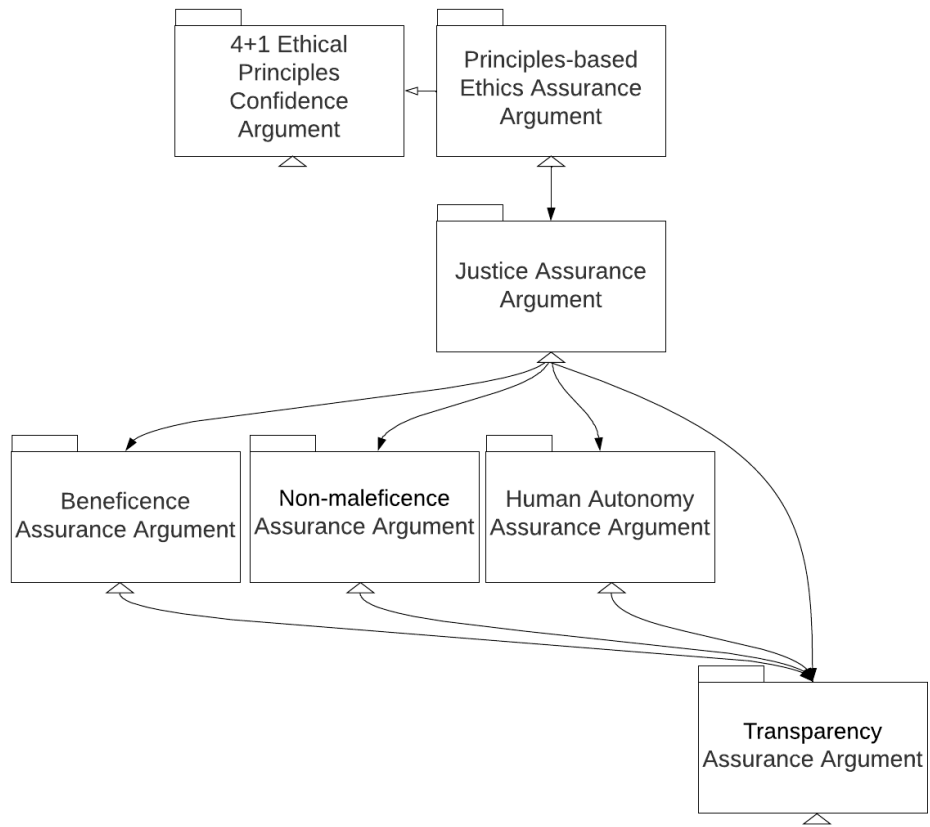

Figure 3: Overview of the Principles-based Ethics Assurance Argument Pattern[1] [41].

The modular structure enables the overall flow of the argument pattern to be shown before its component modules are 'opened up' and described in more detail [41].

In contrast to other approaches, this argument pattern provides a framework for structured, systematic reasoning about these normative goals/ethical desiderata, which also supports holistic deliberation about the trade-offs between them within and across stakeholder groups. To be clear, this argument pattern is not merely a series of rules that one ought to follow, or checkboxes that ought to be crossed off. Rather, it is a reusable template that facilitates reasoning about a particular issue at an abstract level [41].

Indeed this argument pattern, because it is inspired by the assurance case methodology often used for safety assurance cases, has significant advantages in the context of ethical acceptability. The assurance case methodology enables scrutiny, debate and continued improvement via its explicitness, which also facilitates understanding without requiring specialist knowledge, all of which are key to solutions in the complex arena of ethical AI [41]. This framework also allows for the integration and consolidation of multiple evidence sources that will be required to substantiate claims about ethical acceptability [41].

---

[1] Modularity was introduced into GSN in order to support a compositional approach to reasoning about complex systems [25]. It enables the overall flow of the argument pattern to be shown before its components are 'opened up' and explained in more detail.

A summary of the argument is as follows. The top-level goal is contained in the module at the top, titled 'Principles-based Ethics Assurance Argument'.[2] This top-level goal is that, for the intended purpose, the use of the AI-enabled system will be ethically acceptable in the intended context. What constitutes 'ethically acceptable' is based on the notion of a social contract: the system will be ethically acceptable if none of the stakeholders affected by its use could reasonably object to its use in the intended context. The motivation for taking this approach, which has its roots in the work of T. M. Scanlon and John Rawls, is grounded in a commitment to equal respect for all affected stakeholder groups [42,46]. Autonomy, the capacity to live and act according to one's own reasons and motives, plays a key role here since the basic idea is that rational agreement amongst autonomous individuals with equal moral status is what provides the justification for the decision [41]. The idea is that, if the distribution of benefits, tolerable residual risks and tolerable constraints on autonomy is equitable – taking into account existing asymmetries between stakeholder groups and aiming to treat people fairly in light of them – no rational autonomous stakeholder could *reasonably* reject the decision to use the system, and hence we can claim that its use would be ethically acceptable [41]. By 'reasonably reject,' we mean on the assumption that they are not merely seeking some kind of advantage to themselves but are also aimed at finding a conclusion that other autonomous stakeholders, similarly motivated, could not reject [46].

The principles-based structure of the argument pattern is intended to provide a framework to achieve this goal of ethical acceptability. The principle of justice is 'first amongst equals'. The top-level goal of ethical acceptability is immediately supported by the sub-goal that there is an equitable distribution of benefit, tolerable risks of harm and tolerable constraint on human autonomy across affected stakeholders from the use of the system. This sub-goal is contained within the module titled 'Justice Assurance Argument.'

The modules titled 'Beneficence Assurance Argument,' 'Non-maleficence Assurance Argument,' and 'Human Autonomy Assurance Argument' each contain sub-goals about actualising benefits, controlling risks of harm, and managing undue constraints on human autonomy, respectively. Crucially, this information is documented in matrices – a benefits matrix, a risk matrix, and a constraint on autonomy matrix – which furnish the information that is required for reasoning about equitable distributions in the 'Justice Assurance Argument'. The 'Transparency Assurance Argument' plays a vital supporting role by ensuring that there is sufficient visibility and high-quality evidence to have confidence in the claims being made.

## 4 APPLYING THE ASSURANCE ARGUMENT TO DORA

### 4.1 Introduction to the case study

This paper gives the early findings of the first case study applying the principles-based ethics assurance argument pattern to a real-world AI-enabled system. The multi-disciplinary team of authors have applied the assurance argument to Dora over the course of four workshops. Each workshop covered different modules of the argument pattern as set out in Figure 3. The workshops followed the format of semi-structured discussions, where the structure was provided by the detailed decomposition of the argument pattern in [41]. The sections below describe the findings and insights yielded by the semi-structured discussions to date. To note, these are interim results; further workshops with a wider range of stakeholders are being planned.[3] Additionally, it is important to mention that one could apply this ethics assurance argument pattern at several stages in the lifecycle of AI development and deployment. This reasoning would be carried out and communicated pre-deployment of an AI system in its intended context, on the basis of some understanding of its effects during trials, and it would also be reviewed post-deployment to evaluate and address its actual effects.

---

[2] The '4+1 Ethical Principles Confidence Argument' module to the left at the top gives the rationale for structuring the argument according to these four ethical principles, with a supporting principle of transparency.

### 4.2 Interim Results: Overview

Table 1 below gives an overview of a selection of key interim results and insights by stakeholder groups. Table 1 is a high-level presentation of only a few selected results. The full description is given in sections 4.3 - 4.5.

To clarify, considerations having to do with autonomy (row three of Table 1) extend only to a subset of stakeholders, specifically those risk-bearers who are most directly in contact with the system in operation. For Dora, this is patients and clinicians, not the healthcare provider or developer, hence the 'not applicable' result to the latter in the autonomy row. Further, the empty cells for healthcare providers and the developer in the justice row do not indicate that there are no potential inequalities for either stakeholder. Rather, they indicate that none were identified in the relevant workshop. These interim results may be revised as the case study evolves.

Table 1: High-level summary of selected interim results from instantiating the principles-based ethics assurance argument pattern with Dora.

| Ethical principle | Description of the ethical principle | SELECTED interim results by stakeholder group (note this is not the full list) | | | | Insights and priorities for evaluation |
|---|---|---|---|---|---|---|
| | | Patients | Clinicians | Healthcare Providers | Developer | |
| **Beneficence** | The use of Dora should bring benefit to affected stakeholders | Benefit: Physical health and convenience | Benefit: Opportunity to work at the top of their licence | Benefit: Allows delivery of care in constrained contexts | Benefit: Financial and reputational benefits | Some practical and physical benefits are very well substantiated (e.g. cost-effectiveness and patient's practical benefits). Other benefits, such as benefits to clinicians, are presently a little more uncertain and contingent on systemic/structural factors. |
| **Non-maleficence** | The use of Dora should not cause unjustified harm to affected stakeholders | Potential risk: Harm from algorithmic bias | Potential risk: To undermine professional competence | Potential risk: Integration complexities (across different IT systems) | Potential risk: Legal risks and financial risks | The identified risks are appropriately mitigated – but risks to patients, providers and developers are more extensively considered than potential risks to clinicians. |
| **Autonomy** | The use of Dora should not unduly constrain the autonomy of patients and clinicians | Potential constraint: Dora may 'nudge' patient behaviour | Potential constraint: Limited understanding of the system and limited opportunity to consent to its use, as this is a decision made by providers | Not applicable | Not applicable | Within the context of modern clinical practice, patient autonomy and clinician autonomy is more limited than might be expected. In this human context, the impact on autonomy from the use of Dora may not be disproportionate, but this requires further consideration. |

---

[3] Ethics approval has been granted for recording the views of workshop participants concerning the use of Dora.

| | | | | | | |
|---|---|---|---|---|---|---|
| **Justice** | The distribution of benefit, tolerable residual risks and tolerable constraints on autonomy from the use of Dora should be equitable across affected stakeholders | Potential inequity: May disadvantage patient cohorts who require additional support (although may also increase accessibility to care for some patient cohorts) | Potential inequity: May disproportionately constrain clinician autonomy (although may help to share best practice amongst clinicians). Frontline clinicians may also bear the burden of liability/responsibility. | | | Though risks (and benefits) were identified for healthcare providers and the developer, for greater equity we need greater support to patient subgroups and frontline clinicians. |

### 4.3 Beneficence and Non-Maleficence: interim results and insights

The principle of beneficence in the argument pattern requires that use of the system brings benefit to affected stakeholders. While beneficence is central to many of the sets of ethical principles for AI [11,18,37], demonstrating that a system will bring benefit in practice is not in general required by emerging regulation, even though it is implicit in health economics and expected by a body such as The National Institute for Health and Care Excellence (NICE). The principles-based ethics assurance argument seeks to make beneficence more explicit, in order that the good reasons for deploying the AI system are made clear, as is who stands to benefit from its use.

Assessment of the Dora platform revealed that all of the affected stakeholders stand to benefit in various ways from its use. Identified and anticipated benefits to patients were grouped into four kinds: (1) physical health benefits, because Dora delivers a reliable, comprehensive, consistent and safe service to patients, and facilitates continuity of care; (2) benefits to psychological well-being, for example a reduction in emotional harm amongst some patients who noted that Dora eliminates human prejudice from their clinical interaction [14,40]; (3) practical benefits, because Dora is accessible and convenient; and (4) financial benefits, because there is no associated travel cost for patients.

Many of these are likely to materialise. For example, that practical benefits will be realised for patients is well-supported by a user acceptability study by Khavandi et al. [27]. The authors note that patients "appreciated the convenience, availability, and accessibility of Dora" and that 56% of patients surveyed were 'promoters' of the system, i.e., when asked how likely, on a scale of 1-10, they were to recommend the system, they responded with 9 or 10 [27].

Another interesting observation are the impacts on individual patients who are not suitable for a Dora call (for example, if unable to speak English or cognitively impaired). In one pathway with Dora for cataract follow-up, the authors observed the average follow-up time reduced from 6 weeks to a target of 3-4 weeks following implementation – meaning that although individual patients who aren't able to have the calls are disadvantaged, they potentially still benefit from a systemic increase in clinical capacity.

Identified and anticipated benefits to clinicians were: (1) contribution to professional competence, because use of Dora can allow clinicians more opportunity to work at the top of their licence, have better oversight of patient outcomes, allow access to improved patient data, allow the sharing of best practice and enable research opportunities to be involved with pioneering technology; and (2) contribution to psychological well-being, because use of Dora can reduce repetitive, routine tasks and, potentially, reduce burnout. As we highlight in the 'insights' section below, there is relative uncertainty here about the realisation of benefits for this stakeholder group.

Identified benefits to healthcare providers were: (1) contribution to organisational competence, because use of Dora can allow for better workforce management and standardisation of best practice - being relatively free of acute workforce constraints also allows providers a degree of clinical capacity towards delivering consistent, basic care; (2) financial benefits, because use of Dora can allow providers to save costs; and (3) reputational benefits, because use of Dora can allow providers to hit performance targets, improve patient experience and receive hospital kudos and other research opportunities.

Here, too, there is a reasonable expectation that many of these benefits will materialise. Use of Dora is also likely to result in organisational and financial benefits for healthcare providers, for example Bajre and Hart [2] calculate that, when comparing 97 standard post-operative patient follow-ups with 92 Dora post-operative patient follow ups, the cost savings amount to approximately £2,550. Moreover, the authors state that because Dora "may help with reduction in clinicians performing the face-to-face follow-up calls" this "would allow staff resource to be allocated to perform other tasks within the ophthalmology setting" [2].

Identified benefits to the developer were: (1) financial benefits, because Dora meets a demand within the healthcare system and can generate profit; and (2) reputational benefits, because Dora is one of the first products of its kind on the market and part of a sustainable financial model.

The principle of non-maleficence requires that the use of the system does not cause unjustified harm. Physical harm is the traditional concern of safety engineering and safety assurance, and is the central concern of patient safety. But the introduction of data-intensive, increasingly autonomous AI-based systems introduces risks of an extended range of harm, including: psychological harm; misuse of personal data and privacy invasions; harms as a consequence of algorithmic bias against demographic groups; and environmental damage. For reasons of scope, environmental damage was not considered, but this is noted for future instantiations of the argument pattern. Risks such as financial and legal risks were, however, considered in the case study.

Assessment of the Dora platform revealed that all of the affected stakeholders bear various kinds of risks. Identified potential risks to patients were grouped into four kinds: (1) physical health risks, for example if there are false negatives or patient misidentification; (2) risks to psychological well-being, which is a plausible expectation from the fact that fully expressing oneself to Dora is not always possible; (3) risk of bias/discrimination [30], which may occur because Dora's natural language processing is based on a pre-trained language model and because there is an element of human judgement when deciding which patients use Dora; and (4) data security risks, because sensitive patient information is shared.

Many of these risks to patients have been sufficiently managed. The generation of safety evidence through the completion of an FMEA (Failure Modes and Effects Analysis) by the developer, for example, helped to ensure that risks to patient physical health and their sensitive healthcare data are managed. Healthcare providers must also comply with different standards to ensure that risks to patient physical health are managed. In the case of Dora, the DCB0129/0160 standards are of particular importance because they set out context-specific requirements to ensure the effective application of clinical risk management for the deployment, use and maintenance of health information technology systems within the health and care environment (NHS Digital 2018).

Identified potential risks to clinicians were grouped into three kinds: (1) risks to professional competence, which may arise from the fact that use of Dora may deprive trainees from developing their skills by making routine follow-up calls with patients; (2) risks to psychological well-being, because clinicians may see only the difficult cases and consequently could burn out quicker if Dora handles all of the easy non-complicated cases; and (3) legal risks, because the degree to which a clinician is liable for either failing to conform to Dora's recommendation when they should or conforming to Dora's recommendation when they should not, is unclear [29]. As identified in the insights section below, and picked up in the autonomy and justice stages of the case study, these are questions around which there is substantial uncertainty.

Identified potential risks to healthcare providers were: (1) integration complexity risks, which is a plausible expectation given that Dora may over-refer patients and because sensitive patient information is shared between different systems; (2) legal risks, because automated and AI-based systems are novel technologies and regulated via different patchworks of legislation; (3) financial risks; and (4) reputational risks, for example if comments/complaints surface from a vocal minority, even if incidents occur infrequently.

Identified potential risks to the developer were: (1) legal risks, because the regulatory landscape for AI-based technologies is in flux and ill-defined at present; (2) financial risks, because Dora may not be profitable; (3) reputational risks, because the negative experiences of a vocal minority could rapidly proliferate; and (4) security risks, because humans-in-the-loop may make mistakes and because sensitive patient information is shared.

*4.3.1 Insights*

The work is evolving and more data is required to reduce uncertainty as to whether all identified benefits will materialise and whether all identified risks have been sufficiently managed. It is unclear whether, for example, benefits to clinicians will materialise. Use of Dora has the potential to free up clinician time that could presumably be spent on professional development. The reduction of routine, repetitive and inefficient work processes may also alleviate clinician burnout, and so use of Dora may enhance clinician psychological well-being. But these benefits are largely contingent upon other factors in the workplace.

It is unclear whether risks to clinician psychological well-being have been managed and if there is any residual risk to clinicians. As with benefits to clinicians, context appears to play a pivotal role in mitigating risks to psychological well-being (e.g., personality factors, organisational factors, social support, interests outside of medicine, etc.) as does specialty [32]. Further complicating matters is the fact that the use of health information technology can in some contexts increase clinician burnout, but in other contexts help to mitigate burnout [52]. What is clear is that these questions provide developers with a useful guide towards active investigation and evaluation at a deeper, more nuanced level.

**4.4 Autonomy: interim results and insights**

'Autonomy' (of human beings and not machines) concerns people's capacity to control their own destiny to some degree [43], and to live and act according to their own reasons and motives [51].

Within the assurance argument, a principle of respect for human autonomy is broken down into five sub-goals: (1) that the use of Dora should not unduly nudge people into behaviours they would not rationally endorse; (2) that the use of Dora should not be deceptive or misinform people; (3) that Dora makes its decisions on the basis of features or facts in the world that people would consider to be salient; (4) that people should be able to give their informed consent to the use of Dora; and (5) that people should be able physically to intervene in the use of Dora if required, for example if it is malfunctioning.

To note, only the autonomy of immediate risk-bearers from use of the system is considered within the assurance argument, since these are the people who are most immediately subject to harm from the system and therefore most require the capacity to exercise control over it or its influence in their lives. For Dora, this subset of risk-bearers (called 'autonomy risk-bearers') comprises two stakeholder groups: patients and (frontline) clinicians. Also note that the focus is on constraints to autonomy, and not benefits arising from increases in autonomy, which would be covered in the beneficence argument module.

After deliberation within the multi-disciplinary team, the following conclusions concerning the autonomy risk-bearers were reached. Use of Dora does not deceive or misinform people (2). It does, however, to some degree nudge patients (1), since they may adjust their manner of speaking during the call, and this seems to vary across different patient sub-groups. It is uncertain whether clinicians are nudged by the system (1), for example if its use affects how they interpret patient data. Dora does make its recommendations on the basis of features or facts in the world that people would rightly consider to be salient (3), since it has been developed on the basis of the relevant clinical features. Patient capacity to give informed consent to the use of Dora is limited (4), although this is implicit in their accepting and continuing the Dora phone call. Individual clinicians are not generally able to give informed consent to the use of Dora (though providers agree to use it on the whole). This also raises issues around whether there is widespread understanding of the Dora system amongst clinicians, which would contribute to informing their implicit consent to its use. Lastly, both patients and clinicians are able to intervene in the use of Dora (5), in the sense that patients can discontinue the call and clinicians have this capacity at the start of the process, but this capacity decreases for clinicians once the process is underway, i.e., once Dora is given a list of patients to contact. Between points (4) and (5), we noticed an inherent trade-off between clinician autonomy (i.e. having the ability to 'intervene' in real-time on the call), and realising the full benefits of an autonomous system (for example, having 80 patients on a clinic list being simultaneously called at the same time).

*4.4.1 Insights*

Human autonomy is highly context-sensitive. Analysis of the use of Dora revealed that patients and clinicians possess less autonomy in a healthcare setting than one might initially suspect. But it was noted that these restrictions on human autonomy are not entirely outside the norm. Patients, for example, often have the chance to opt-out but not always opt-in to specific healthcare treatments or pathways, and this holds for the use of Dora which is why patient capacity for giving informed consent and intervening is minimal (sub-goals 4 and 5). Given this context it is not, perhaps, an undue constraint on autonomy.

This brings an important question to the fore: what exactly in the healthcare setting constitutes an undue constraint on autonomy? Like patients, the capacity of individual clinicians to give informed consent specifically to the use of Dora is relatively constrained. Yet, especially within the context of a public health system like the NHS, protocols and procedures are often decided at a departmental, regional or even national level – and individual clinicians already have only limited capacity to 'consent' to the use of these in their individual practice. In this context, whether a system like Dora *unduly* constrains clinician autonomy remains an open question – and one which can be explored further as the case study evolves with the participation of a wider group of stakeholders in future workshops.

One possible impact of the identified constraints on human autonomy is that it risks making those closest to the system – and particularly, in this case, the frontline clinician – a 'moral crumple zone' [9], whereby they may in practice bear responsibility for harmful outcomes involving an AI despite having limited control over its use or behaviour. This prompts the observation that adjustments to increase or maintain clinician autonomy should be identified, and the lines of responsibility and accountability for the consequences of using Dora need to be made explicit by providers as well as regulators. This is picked up in the discussion below.

**4.5 Interim Results: Justice**

Prior to the justice workshop, we considered the assumptions and reasons underlying claims made in the previous workshops, e.g., that because Dora is consistent, this will lead to better patient health outcomes. This was the purpose of the transparency workshop, where 'transparency' refers broadly to the "visibility" of information. Implicit assumptions underlying claims made in the previous workshops were therefore made explicit for reasoning about. Where possible, evidence was also produced to support claims made in earlier workshops (e.g., user acceptability study, economic analysis report, sustainability report, etc.). Transparency is intimately connected to the entire assurance argument (beneficence, non-maleficence, human autonomy and justice). This is because making the implicit explicit helps to justify confidence in the claims made throughout, and ultimately it helps to ensure that what is reasoned about at the level of justice is not merely conjecture.

In the principles-based ethics assurance argument pattern, the principle of justice requires that the distribution of benefit, tolerable residual risks of harm and tolerable constraints on human autonomy be equitable across all affected stakeholders. This reasoning underscores the social contract approach. The idea is that, if this distribution is equitable, no affected stakeholders could reasonably object to the use of Dora.

The first stage of the justice workshop was to consider whether the use of Dora incurs any ethically problematic role combinations. This part of the argument derives from the ethical risk analysis work of Hansson [21]. Problematic role combinations are broken down as follows: (1) there should be no risk-bearer who bears only risk from Dora; (2) no risk-bearer who only receives minimal benefit should be left uncompensated for that risk; (3) no immediate risk-bearer should have their autonomy unduly constrained. If problematic role combinations cannot be eliminated, it would not be ethically acceptable to deploy the described system in the intended context.

The initial findings were that ethically problematic role combinations do not clearly arise from the use of Dora, on the assumption that the analysis in the previous stages of the case study is correct – and assuming that, within the context, the constraints on autonomy do not count as 'undue' or 'intolerable'. As such, this discussion also raised questions for consideration about whether the identified constraints on patient and frontline clinician autonomy (i.e., the manner or degree to which Dora 'nudges' behaviour, and the limited options for 'opting out' of the use of Dora) need to be managed or adjusted in the clinical pathway.

The next stage was to consider whether use of Dora entrenches existing inequalities amongst stakeholder groups. This prompted the observation that use of Dora may entrench some existing inequalities against frontline clinicians. Given the inability of clinicians to give their informed consent to the use of systems such as Dora, coupled with the fact that the use of AI-based systems complicates and obscures ascriptions of accountability [36], there is a danger that clinicians will be held responsible (hence the potential 'moral crumple zone') for the consequences of using systems they had limited capacity to consent to use in the first place. This consideration needs to be balanced against the fact that, in the presence of the right conditions, use of Dora could also facilitate increased clinician autonomy by allowing them to work at the top of their licence as well as pursue in-depth training and skills. Use of Dora may also entrench existing inequalities against sub-cohorts of patients. Patients whose native language is not English, who have more nuanced healthcare needs, and patients less comfortable on the telephone continue to be marginalised and may perceive that they are receiving, and may in fact receive, substandard healthcare as a result of use of Dora.

The third stage was to conduct a 'reflective equilibrium' decision procedure about the distribution of benefits, tolerable residual risks and tolerable autonomy constraints across affected stakeholders – and to consider what adjustments could be made at stages of the product development and deployment lifecycle to make this distribution fair. Reflective equilibrium is inspired by the work of the political philosopher John Rawls [42]. It is the end-point of a decision-procedure which is reached when people work backwards and forwards between their judgements about a state of affairs and ethical principles (as well as non-moral considerations, such as practical or legal facts), and make adjustments as appropriate, until an acceptable coherence of opinion is reached [6,42]. Reflective equilibrium is achieved when none of the people involved are inclined to make any further adjustments in order to accept the state of affairs in question. One approach is to apply a hypothetical 'veil of ignorance' when engaging in this decision procedure, whereby participants do not know their position in society [42] – or, in the Dora case study, which stakeholder group they belong to. Another approach is to require everyone to focus on the same goal – in this case, the goal of ethical acceptability across all affected stakeholders – and to use this as a reference point for what it would be reasonable to agree to and what adjustments should be made. Because in practice it may be unrealistic to expect people to reason from behind a 'veil of ignorance,' the ethics assurance argument takes the latter approach; but this may change in later iterations of the argument pattern.

*4.5.1 Insights*

The following three themes emerged from the reflective equilibrium decision procedure.

Theme 1: Certain sub-cohorts of patients require additional support. While most patients could be expected not to have a reason to object to the use of Dora, certain sub-cohorts of patients, such as those with more nuanced healthcare needs or those that are socially isolated, may require additional protection. Suggested adjustments to accommodate these patients included giving patients the choice to have a human call instead of Dora and/or to have a contact number to call when they have concerns. Healthcare providers and/or the developer could also reinvest some of their gains in people and in infrastructure.

Theme 2: Clinicians may bear disproportionately more risk. The second theme extracted touched on issues mentioned above, around the autonomy of frontline clinicians, uncertainties about the risks to them and uncertainties about the realisation of benefits for them. Suggested adjustments included encouraging providers to reduce the risk to the frontline clinician by clarifying questions around responsibility and accountability, ensuring that they are both adequately supported and are, for example, given opportunities to improve professional competence to ensure that they reap benefits from use of the system. It was also suggested that healthcare providers and/or the developer could advocate for advances in legislation to clarify the clinician's responsibilities when AI is being used in the pathway.

Theme 3: The third theme highlighted was that there are many systemic/structural factors that may need adjusting before the distribution of benefits, tolerable residual risk and constraints on human autonomy is such that none of the affected stakeholders could reasonably object to the use of Dora. Challenges include incorporating appropriate non-technical "opt-out" options as part of the design of the system. Suggested adjustments included having providers change how information is communicated and advocating for a clear legal framework so that both clinicians and patients clearly understand how the use of Dora affects them.

An overall insight from the workshops, which came out strongly in the justice workshop, was that the principles-based ethics assurance framework surfaced issues that might otherwise have been overlooked or not strongly emphasised. Though challenging to apply, it prompted a candid, integrated approach to reasoning about the ethical acceptability of the use of Dora which seems initially promising as a framework for securing deep, well-justified acceptance of AI systems in complex, and often constrained, real-world socio-technical settings.

**4.6 Reflections from a clinical point of view**

As AI systems such as Dora, which can operate independently of direct and continuous human intervention, become more prevalent in healthcare, clinicians will increasingly need to evaluate their value, risks, and applicability to their practice, similar to how they now routinely evaluate pharmaceutical agents or medical devices. However, modern clinical training does not equip clinicians with a practical framework for first-principles reasoning around the use of an autonomous system in their practice.

Consider this thought experiment: if a hypothetical diagnostic AI system that was perfectly efficacious, with robust long-term follow-up data, was shown to be cost-effective with a full health economic analysis, and if its algorithms were explainable, transparent, and unbiased – would this system be acceptable for wider deployment? Unlike a pharmaceutical agent, an AI-enabled system capable of influencing – or outright making – decisions, fundamentally changes the dynamics of care [20]. In isolation, considerations like safety, health economics or model performance, whilst all individually crucial, do not add up to form a holistic assurance-case for deployment of a system.

The principles-based ethics assurance approach is a powerful framework that enables us to have this appropriately holistic view of the complex and multifaceted nature of increasingly autonomous AI-based systems deployed in a clinical setting. It allows clinicians and stakeholders to balance and weigh important issues around concerns such as bias [38], algorithmic explainability [44], and accountability [20], within a systematic framework already familiar to clinical practitioners.

AI-based medical devices - like any clinical intervention - are often deployed in highly specific contexts and patient populations. The principles-based ethics assurance argument pattern helps us demonstrate a process of structured reasoning which forms the basis for flexibly evaluating the ethical implications of using such systems in their intended contexts of use. This is a uniquely human-centred approach and, by placing the emphasis around the individuals using an AI system in a real-world context, it allows us to surface disproportionate harms to individuals. It also enables an appraisal of evidence appropriate to the current stage of development and deployment, rather than relying on an individual model, method, or disease-specific framework.

Finally, it allows a degree of transparency and consistency, ensuring all relevant issues are considered based on a shared understanding of what constitutes ethical behaviours. The importance of balancing risks and evaluating the potential benefits and drawbacks of autonomous clinical AI systems is reflective of the real-world decision-making processes that healthcare professionals already engage in on a daily basis.

## 5 CONCLUSIONS AND FUTURE WORK

### 5.1 Interim conclusions

Applying the principles-based ethics assurance argument pattern to Dora has revealed the positive ethical impacts of the platform (it is safe, patients appreciate the convenience of it, and it is delivering care in the highly constrained context of a stretched healthcare system). The case study to date has also revealed areas to prioritise for evaluation, such as consideration of the systemic factors upon which potential benefits are contingent, and more consideration of the risk borne by clinicians.

A key factor is that it offers a framework for integrated or holistic reasoning about the ethical acceptability of using an AI system, which accords equal status to all affected stakeholder groups. It is grounded in notions of distributive equity and a social contract: it is not just concerned with showing an aggregation of benefit and risk reduction, nor is it limited to those that can be relatively easily measured. This has illuminated new insights that, if addressed, could have significant advantages for identified stakeholders and ensure that stakeholder subgroups are not disproportionately disadvantaged. One example is around the autonomy of clinicians, which may otherwise have been overlooked. While work remains to be done on what constitutes an 'undue' constraint on autonomy in the existing context, addressing this could massively impact acceptance of – and ethical acceptability of – the system when deployed at scale.

The justice workshop in particular revealed that the argument pattern offers a surprisingly practical framework. Working through the reflective equilibrium process meant that participants did not just identify problems, but started to construct solutions, in the form of adjustments that could reduce ethical disparities and stop inequalities from becoming entrenched. The identification of these practical solutions suggests that much will be gained from including a wider range of stakeholders in the reflective equilibrium procedure to draw on their experience and insights, and this is planned for the next step of the case study. At the same time, though the framework is challenging and sets a high and ambitious threshold for ethical acceptability, it also enables us to be realistic about real-world constraints – such as the constraints faced by a public healthcare system in which demand is increasing beyond the capacity of healthcare staff – and the trade-offs that need to be made in these circumstances. The point is about making these trade-offs fairly and protecting those who might otherwise bear the burden of risk.

### 5.2 Limitations and future work

In section 2.2, we said that working through a principles-based ethics assurance argument can be seen as an example of RRI in action. It goes beyond regulatory compliance and provides a framework for deep reflection on the ethical impact of defined AI technologies, involving dialogue between different stakeholders and facilitating trustworthy AI development and deployment.

We can also apply RRI to our own case study activities at this interim stage, and – as we consider the next steps – evaluate them against the four central dimensions of RRI: anticipation, reflexivity, inclusion and responsiveness [50]. The process has exemplified the 'anticipation' and 'reflexivity' dimensions. It has enabled us to *anticipate* previously unidentified possible outcomes and it has been a *reflexive* process, because we have held a mirror up to assumptions, such as the assumption that a reduction in one kind of routine task will naturally benefit clinicians. It has also created an awareness of the boundaries of current relevant knowledge which can and should be explored, for example comparisons with the safety of human-made calls in this pathway and how to measure psychological impact, and how to evaluate informed consent. Important next steps are to *include* a wider range of stakeholders in the discussion, including patient representatives, frontline clinicians, managers, the technical team and regulators. The imperative here will be to be responsive to these stakeholders' values, priorities and circumstances in the development of the framework, in this and future instantiations of the argument pattern, and in reporting.

Further future work is to use the interim findings to guide and prioritise specific, prospective evaluation metrics. An ethical, principles-based approach to designing clinical evaluation for an autonomous system has previously been proposed [1], but efforts thus far have not factored in findings unearthed during the process of applying the principles-based argument pattern. Our hypothesis is that the findings help identify such gaps and also provide a practical and ethical basis for developers and users to decide on key areas of evaluation to prioritise.

Preliminary results of this case study also reveal that there is important work to be done concerning the explication of "unjustified" harm and "undue" or "intolerable" constraints on autonomy in the healthcare context and in other contexts of interest. Facilitating dialogue between affected stakeholders is therefore critically important to ensure that all parties understand the effects that deploying a system like Dora may have. Moreover, those effects ought to be monitored over the lifecycle of the use of the system so that sustainable trust and acceptability can be fostered.

Relatedly, future work would tackle the problem of incommensurability explicitly. It is, in short, non-trivial to compare different benefits (e.g., physical health benefits and professional competence benefits), let alone compare benefits with risks or with constraints on autonomy. Even granting that appropriate metrics for different benefits, risks and constraints on autonomy exist, comparing them will remain a challenge given their incommensurability, i.e., their inability to be ranked on some common cardinal scale like dollars or QALYs (quality-adjusted life years). How such comparisons ought to proceed is an important question that should involve input from all affected stakeholders.

**ACKNOWLEDGEMENTS**

We would like to thank James Godwin and Nick de Pennington for their insight and early involvement in the workshops. We would also like to thank John McDermid for his role in developing the principles-based ethics assurance argument pattern. This work was supported by the Assuring Autonomy International Programme, a partnership between Lloyd's Register Foundation and the University of York and by the Engineering and Physical Sciences Research Council through the Assuring Responsibility for Trustworthy Autonomous Systems project (EP/W011239/1).

**REFERENCES**

[1] Michael D. Abràmoff, Danny Tobey, and Danton S. Char. 2020. Lessons Learned About Autonomous AI: Finding a Safe, Efficacious, and Ethical Path Through the Development Process. Am. J. Ophthalmol. 214, (June 2020), 134–142. DOI:https://doi.org/10.1016/j.ajo.2020.02.022

[2] Mamta Bajre and Julie Hart. 2022. Economic Analysis Report: AI_AWARD01852 Automated Autonomous Telemedicine - Cataract Surgery Follow-up at Two NHS Trusts.

[3] T Beauchamp and J Childress. 1979. Principles of Biomedical Ethics. Oxford University Press, New York.

[4] Stan Benjamens, Pranavsingh Dhunnoo, and Bertalan Meskó. 2020. The state of artificial intelligence-based FDA-approved medical devices and algorithms: an online database. NPJ Digit. Med. 3, (2020), 118. DOI:https://doi.org/10.1038/s41746-020-00324-0

[5] Christopher Burr and David Leslie. 2022. Ethical assurance: a practical approach to the responsible design, development, and deployment of data-driven technologies. AI Ethics 3, 1 (February 2022), 73–98. DOI:https://doi.org/10.1007/s43681-022-00178-0

[6] Norman Daniels. 2020. Reflective Equilibrium. In The Stanford Encyclopedia of Philosophy (Summer 2020), Edward N. Zalta (ed.). Metaphysics Research Lab, Stanford University. Retrieved March 7, 2023 from https://plato.stanford.edu/archives/sum2020/entries/reflective-equilibrium/

[7] Irene Dankwa-Mullan, Elisabeth Lee Scheufele, Michael E. Matheny, Yuri Quintana, Wendy W. Chapman, Gretchen Jackson, and Brett R. South. 2021. A Proposed Framework on Integrating Health Equity and Racial Justice into the Artificial Intelligence Development Lifecycle. J. Health Care Poor Underserved 32, 2 (2021), 300–317. DOI:https://doi.org/10.1353/hpu.2021.0065

[8] Jeffrey De Fauw, Joseph R. Ledsam, Bernardino Romera-Paredes, Stanislav Nikolov, Nenad Tomasev, Sam Blackwell, Harry Askham, Xavier Glorot, Brendan O'Donoghue, Daniel Visentin, George van den Driessche, Balaji Lakshminarayanan, Clemens Meyer, Faith Mackinder, Simon Bouton, Kareem Ayoub, Reena Chopra, Dominic King, Alan Karthikesalingam, Cían O. Hughes, Rosalind Raine, Julian Hughes, Dawn A. Sim, Catherine Egan, Adnan Tufail, Hugh Montgomery, Demis Hassabis, Geraint Rees, Trevor Back, Peng T. Khaw, Mustafa Suleyman, Julien Cornebise, Pearse A. Keane, and Olaf Ronneberger. 2018. Clinically applicable deep learning for diagnosis and referral in retinal disease. Nat. Med. 24, 9 (September 2018), 1342–1350. DOI:https://doi.org/10.1038/s41591-018-0107-6

[9] Madeleine Clare Elish. 2019. Moral Crumple Zones: Cautionary Tales in Human-Robot Interaction. Engag. Sci. Technol. Soc. 5, (March 2019), 40–60. DOI:https://doi.org/10.17351/ests2019.260

[10] Andre Esteva, Brett Kuprel, Roberto A. Novoa, Justin Ko, Susan M. Swetter, Helen M. Blau, and Sebastian Thrun. 2017. Dermatologist-level classification of skin cancer with deep neural networks. Nature 542, 7639 (February 2017), 115–118. DOI:https://doi.org/10.1038/nature21056

[11] European Commission and Content and Technology Directorate-General for Communications Networks. 2019. Ethics guidelines for trustworthy AI. Publications Office. DOI:https://doi.org/10.2759/346720

[12] FDA. 2022. Artificial Intelligence and Machine Learning in Software as a Medical Device. FDA (September 2022). Retrieved March 1, 2023 from https://www.fda.gov/medical-devices/software-medical-device-samd/artificial-intelligence-and-machine-learning-software-medical-device

[13] Jean Feng, Rachael V. Phillips, Ivana Malenica, Andrew Bishara, Alan E. Hubbard, Leo A. Celi, and Romain Pirracchio. 2022. Clinical artificial intelligence quality improvement: towards continual monitoring and updating of AI algorithms in healthcare. Npj Digit. Med. 5, 1 (May 2022), 1–9. DOI:https://doi.org/10.1038/s41746-022-00611-y

[14] Chloë FitzGerald and Samia Hurst. 2017. Implicit bias in healthcare professionals: a systematic review. BMC Med. Ethics 18, 1 (March 2017), 19. DOI:https://doi.org/10.1186/s12910-017-0179-8

[15] Jessica Fjeld, Nele Achten, Hannah Hilligoss, Adam Nagy, and Madhulika Srikumar. 2020. Principled Artificial Intelligence: Mapping Consensus in Ethical and Rights-Based Approaches to Principles for AI. DOI:https://doi.org/10.2139/ssrn.3518482

[16] Luciano Floridi and Josh Cowls. 2019. A Unified Framework of Five Principles for AI in Society. Harv. Data Sci. Rev. 1, 1 (July 2019). DOI:https://doi.org/10.1162/99608f92.8cd550d1

[17] Luciano Floridi, Josh Cowls, Monica Beltrametti, Raja Chatila, Patrice Chazerand, Virginia Dignum, Christoph Luetge, Robert Madelin, Ugo Pagallo, Francesca Rossi, Burkhard Schafer, Peggy Valcke, and Effy Vayena. 2018. AI4People—An Ethical Framework for a Good AI Society: Opportunities, Risks, Principles, and Recommendations. Minds Mach. 28, 4 (December 2018), 689–707. DOI:https://doi.org/10.1007/s11023-018-9482-5

[18] Future of Life Institute. 2017. AI Principles. Retrieved March 3, 2023 from https://futureoflife.org/open-letter/ai-principles/

[19] Oliver Willard Gardiner, Mohita Chowdhury, Ernest Lim, Aisling Higham, and Nick de Pennington. 2022. Can deep learning models understand natural language descriptions of patient symptoms following cataract surgery? Invest. Ophthalmol. Vis. Sci. 63, 7 (June 2022), 1694-F0012.

[20] Ibrahim Habli, Tom Lawton, and Zoe Porter. 2020. Artificial intelligence in health care: accountability and safety. Bull. World Health Organ. 98, 4 (April 2020), 251–256. DOI:https://doi.org/10.2471/BLT.19.237487

[21] Sven Ove Hansson. 2018. How to Perform an Ethical Risk Analysis (eRA). Risk Anal. 38, 9 (2018), 1820–1829. DOI:https://doi.org/10.1111/risa.12978

[22] Aisling Higham, Ernest Lim, Filip Tarcoveanu, Christopher King, Sarah Maling, and Mike Adams. 2023. Real world use and efficiency savings for clinical staff when using automated telephone follow up after routine cataract surgery. American Society of Cataract and Refractive Surgeons Conference Abstract 2023 (2023).

[23] Steven Horng, David A. Sontag, Yoni Halpern, Yacine Jernite, Nathan I. Shapiro, and Larry A. Nathanson. 2017. Creating an automated trigger for sepsis clinical decision support at emergency department triage using machine learning. PloS One 12, 4 (2017), e0174708. DOI:https://doi.org/10.1371/journal.pone.0174708

[24] Anna Jobin, Marcello Ienca, and Effy Vayena. 2019. The global landscape of AI ethics guidelines. Nat. Mach. Intell. 1, 9 (September 2019), 389–399. DOI:https://doi.org/10.1038/s42256-019-0088-2

[25] Tim Kelly. 2001. Concepts and principles of compositional safety case construction. Contract Research Report for QinetiQ COMSA/2001/1/1, 34. Contract Research Report for QinetiQ COMSA/2001/1/1 34, (2001). Retrieved from https://www-users.cs.york.ac.uk/~tpk/CompositionalSafetyCases.pdf

[26] Timothy Patrick Kelly. 1998. Arguing Safety – A Systematic Approach to Managing Safety Cases. PhD Thesis Dep. Comput. Sci. Univ. York (1998). Retrieved from https://citeseerx.ist.psu.edu/document?repid=rep1&type=pdf&doi=81d2e41a5673a8d4a0d7c78ca3d0b0ff26165991

[27] Sarah Khavandi, Ernest Lim, Aisling Higham, Nick de Pennington, Mandeep Bindra, Sarah Maling, Mike Adams, and Guy Mole. 2022. User-acceptability of an automated telephone call for post-operative follow-up after uncomplicated cataract surgery. Eye Lond. Engl. (October 2022). DOI:https://doi.org/10.1038/s41433-022-02289-8

[28] Tiffany H. Kung, Morgan Cheatham, Arielle Medenilla, Czarina Sillos, Lorie De Leon, Camille Elepaño, Maria Madriaga, Rimel Aggabao, Giezel Diaz-Candido, James Maningo, and Victor Tseng. 2023. Performance of ChatGPT on USMLE: Potential for AI-assisted medical education using large language models. PLOS Digit. Health 2, 2 (February 2023), e0000198. DOI:https://doi.org/10.1371/journal.pdig.0000198

[29] Tom Lawton, Phillip Moran, Zoe Porter, Alice Cunningham, Nathan Hughes, Ioanna Iacovides, Yan Jia, Vishal Sharma, and Ibrahim Habli. 2023. Clinicians Risk Becoming "Liability Sinks" for Artif. Inell. Preprints.

[30] Xiaoxuan Liu, Ben Glocker, Melissa M. McCradden, Marzyeh Ghassemi, Alastair K. Denniston, and Lauren Oakden-Rayner. 2022. The medical algorithmic audit. Lancet Digit. Health 4, 5 (May 2022), e384–e397. DOI:https://doi.org/10.1016/S2589-7500(22)00003-6

[31] Nina Markl. 2022. Language variation and algorithmic bias: understanding algorithmic bias in British English automatic speech recognition. In 2022 ACM Conference on Fairness, Accountability, and Transparency (FAccT '22), Association for Computing Machinery, New York, NY, USA, 521–534. DOI:https://doi.org/10.1145/3531146.3533117

[32] John A. McDermid. 1994. Support for safety cases and safety arguments using SAM. Reliab. Eng. Syst. Saf. 43, 2 (January 1994), 111–127. DOI:https://doi.org/10.1016/0951-8320(94)90057-4

[33] Nicola McKinley, R. Scott McCain, Liam Convie, Mike Clarke, Martin Dempster, William Jeffrey Campbell, and Stephen James Kirk. 2020. Resilience, burnout and coping mechanisms in UK doctors: a cross-sectional study. BMJ Open 10, 1 (January 2020), e031765. DOI:https://doi.org/10.1136/bmjopen-2019-031765

[34] MHRA. 2022. Software and AI as a Medical Device Change Programme - Roadmap. GOV.UK. Retrieved March 8, 2023 from https://www.gov.uk/government/publications/software-and-ai-as-a-medical-device-change-programme/software-and-ai-as-a-medical-device-change-programme-roadmap

[35] Jessica Morley, Luciano Floridi, Libby Kinsey, and Anat Elhalal. 2020. From What to How: An Initial Review of Publicly Available AI Ethics Tools, Methods and Research to Translate Principles into Practices. Sci. Eng. Ethics 26, 4 (August 2020), 2141–2168. DOI:https://doi.org/10.1007/s11948-019-00165-5

[36] NHS Digital. 2018. DCB0129: Clinical Risk Management: its Application in the Manufacture of Health IT Systems. NHS Digital. Retrieved March 6, 2023 from https://digital.nhs.uk/data-and-information/information-standards/information-standards-and-data-collections-including-extractions/publications-and-notifications/standards-and-collections/dcb0129-clinical-risk-management-its-application-in-the-manufacture-of-health-it-systems

[37] Helen Nissenbaum. 1996. Accountability in a computerized society. Sci. Eng. Ethics 2, 1 (March 1996), 25–42. DOI:https://doi.org/10.1007/BF02639315

[38] OECD Legal Instruments. 2019. Recommendation of the Council on Artificial Intelligence. Retrieved March 3, 2023 from https://legalinstruments.oecd.org/en/instruments/OECD-LEGAL-0449

[39] Trishan Panch, Heather Mattie, and Rifat Atun. 2019. Artificial intelligence and algorithmic bias: implications for health systems. J. Glob. Health 9, 2 (2019), 020318. DOI:https://doi.org/10.7189/jogh.09.020318

[40] Nick de Pennington, Guy Mole, Ernest Lim, Madison Milne-Ives, Eduardo Normando, Kanmin Xue, and Edward Meinert. 2021. Safety and Acceptability of a Natural Language Artificial Intelligence Assistant to Deliver Clinical Follow-up to Cataract Surgery Patients: Proposal. JMIR Res. Protoc. 10, 7 (July 2021), e27227. DOI:https://doi.org/10.2196/27227

[41] Tara S. Peris, Bethany A. Teachman, and Brian A. Nosek. 2008. Implicit and Explicit Stigma of Mental Illness: Links to Clinical Care. J. Nerv. Ment. Dis. 196, 10 (October 2008), 752. DOI:https://doi.org/10.1097/NMD.0b013e3181879dfd

[42] Zoe Porter, Ibrahim Habli, John McDermid, and Marten Kaas. 2023. A Principles-based Ethical Assurance Argument for AI and Autonomous Systems. AI Ethics, forthcoming.

[43] John Rawls. 1971. A Theory of Justice. Harvard University Press, Cambridge, MA.

[44] J Raz. 1986. The morality of freedom. Clarendon Press.

[45] Sandeep Reddy. 2022. Explainability and artificial intelligence in medicine. Lancet Digit. Health 4, 4 (April 2022), e214–e215. DOI:https://doi.org/10.1016/S2589-7500(22)00029-2

[46] Regulatory Horizons Council. 2021. Regulatory Horizons Council report on medical devices. Retrieved March 7, 2023 from https://assets.publishing.service.gov.uk/government/uploads/system/uploads/attachment_data/file/1012043/rhc-medical-report.pdf

[47] Thomas Scanlon. 1998. What we owe to each other. Belknap Press of Harvard University Press, Cambridge, Mass.

[48] Karan Singhal, Shekoofeh Azizi, Tao Tu, S. Sara Mahdavi, Jason Wei, Hyung Won Chung, Nathan Scales, Ajay Tanwani, Heather Cole-Lewis, Stephen Pfohl, Perry Payne, Martin Seneviratne, Paul Gamble, Chris Kelly, Nathaneal Scharli, Aakanksha Chowdhery, Philip Mansfield, Blaise Aguera y Arcas, Dale Webster, Greg S. Corrado, Yossi Matias, Katherine Chou, Juraj Gottweis, Nenad Tomasev, Yun Liu, Alvin Rajkomar, Joelle Barral, Christopher Semturs, Alan Karthikesalingam, and Vivek Natarajan. 2022. Large Language Models Encode Clinical Knowledge. DOI:https://doi.org/10.48550/arXiv.2212.13138

[49] Brian Cantwell Smith. 2019. The Promise of Artificial Intelligence. MIT Press. Retrieved March 8, 2023 from https://mitpress.mit.edu/9780262043045/the-promise-of-artificial-intelligence/

[50] Bernd Carsten Stahl, Simisola Akintoye, Lise Bitsch, Berit Bringedal, Damian Eke, Michele Farisco, Karin Grasenick, Manuel Guerrero, William Knight, Tonii Leach, Sven Nyholm, George Ogoh, Achim Rosemann, Arleen Salles, Julia Trattnig, and Inga Ulnicane. 2021. From Responsible Research and Innovation to responsibility by design. J. Responsible Innov. 8, 2 (May 2021), 175–198. DOI:https://doi.org/10.1080/23299460.2021.1955613

[51] Jack Stilgoe, Richard Owen, and Phil Macnaghten. 2013. Developing a framework for responsible innovation. Res. Policy 42, 9 (November 2013), 1568–1580. DOI:https://doi.org/10.1016/j.respol.2013.05.008

[52] The King's Fund. 2022. NHS staffing shortages. The King's Fund. Retrieved January 31, 2023 from https://www.kingsfund.org.uk/publications/nhs-staffing-shortages

[53] Carissa Véliz. 2019. Three things digital ethics can learn from medical ethics. Nat. Electron. 2, 8 (August 2019), 316–318. DOI:https://doi.org/10.1038/s41928-019-0294-2

[54] Danny T. Y. Wu, Catherine Xu, Abraham Kim, Shwetha Bindhu, Kenneth E. Mah, and Mark H. Eckman. 2021. A Scoping Review of Health Information Technology in Clinician Burnout. Appl. Clin. Inform. 12, 3 (May 2021), 597–620. DOI:https://doi.org/10.1055/s-0041-1731399